\documentclass{PoS}

\usepackage{amsmath,bbm}
\usepackage{amsfonts}

\newcommand{\beq}{\begin{equation}}
\newcommand{\eeq}{\end{equation}}
\newcommand{\bea}{\begin{eqnarray}}
\newcommand{\eea}{\end{eqnarray}}
\newcommand{\beas}{\begin{eqnarray*}}
\newcommand{\eeas}{\end{eqnarray*}}
\newcommand{\eq}{\begin{equation}}
\newcommand{\en}{\end{equation}}
\newcommand{\eqa}{\begin{eqnarray}}
\newcommand{\ena}{\end{eqnarray}}

\title{The gluon propagator in Coulomb gauge from the lattice }

\ShortTitle{The gluon propagator in Coulomb gauge from the lattice }

\author{\speaker{Giuseppe Burgio}\\
        E-mail: \email{burgio@tphys.physik.uni-tuebingen.de}}

\author{Markus Quandt}

\author{Hugo Reinhardt\\        
        Institut f\"ur Theoretische Physik\\ 
        Auf der Morgenstelle 14\\ 
        D-72076 T\"ubingen (D)}

\abstract{We show that in the lattice Hamiltonian
limit the static transverse propagator 
$D(|\vec{p}|)\propto\int d p_0 D(|\vec{p}|,p_0)$ satisfies 
multiplicative renormalizability. We give a procedure to calculate 
$D(|\vec{p}|)$ on available
lattices at finite temporal spacing. The result agrees 
at all momenta with the Gribov formula $D(|\vec{p}|)\propto(|\vec{p}|^2+
M^4 |\vec{p}|^{-2})^{-\frac{1}{2}}$, with $M=0.88(1) {\rm GeV} \simeq 2 \sqrt{\sigma}$.}

\FullConference{8th Conference Quark Confinement and the Hadron Spectrum\\
		 September 1-6, 2008\\
		 Mainz. Germany}

\begin{document}

\section{Introduction}
\label{sec:introduction}

The original Gribov-Zwanziger confinement scenario
\cite{Gribov:1977wm,Zwanziger:1995cv} predicts an IR
vanishing static gluon propagator $D(|\vec{p}|)$ in Coulomb gauge. 
The gluon propagator is also at the heart of variational estimates to the 
ground state wave function \cite{Szczepaniak:2001rg,Feuchter:2004mk}.
A cross check of continuum results with lattice calculations had, however, 
long failed. A first study for SU(2) at fixed $\beta=2.2$
indicated compatibility with Gribov's formula in the IR but was inconclusive 
in the UV \cite{Cucchieri:2000gu}. Later studies in SU(2) and SU(3) showed 
for $D(|\vec{p}|)$ strong scaling violations and a UV behaviour at odds with 
simple dimensional arguments \cite{Langfeld:2004qs,Quandt:2007qd,Voigt:2007wd}.
All these works calculate $D(|\vec{p}|)$ fixing the Coulomb gauge only 
at a given time-slice, neglecting the residual gauge freedom of temporal 
links. Also one takes for granted that multiplicative renormalizability 
holds for the full propagator 
$D(|\vec{p}|,p_0)$, although perturbative results point at a 
more complex picture \cite{Watson:2007mz}. We report here the results
first obtained in \cite{Burgio:2008jr}, where a different strategy was 
adopted, fixing analytically the residual gauge and then studying the
renormalization of the full spatial propagator $D(|\vec{p}|,p_0)$. We 
show that the Gribov formula \cite{Gribov:1977wm} perfectly describes the 
lattice data for the static propagator. We refer to our original
paper \cite{Burgio:2008jr} for details about conventions and notations. To 
achieve a good gauge fixing we adapt the algorithms developed in 
\cite{Bogolubsky:2005wf,Bogolubsky:2007bw}.

\section{Results}
\label{sec:results}

\begin{figure}
\includegraphics{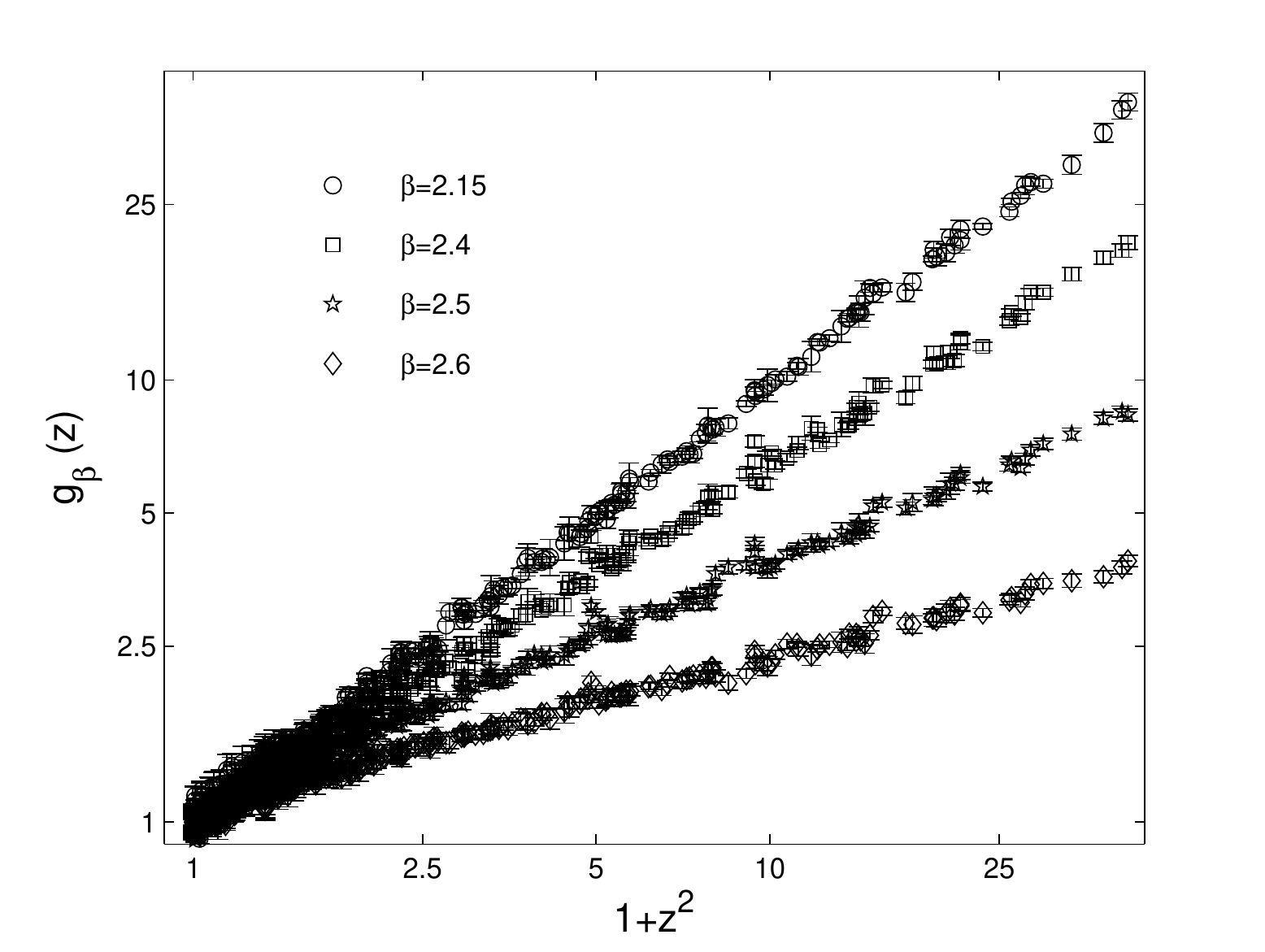}
\caption{Data for $g_\beta(z)$ vs $1+z^2$ in log-log scale, $L=24$. For sake of readability not all $\beta$ are shown.}
\label{fig:polaw}
\end{figure}

\begin{figure}
\includegraphics{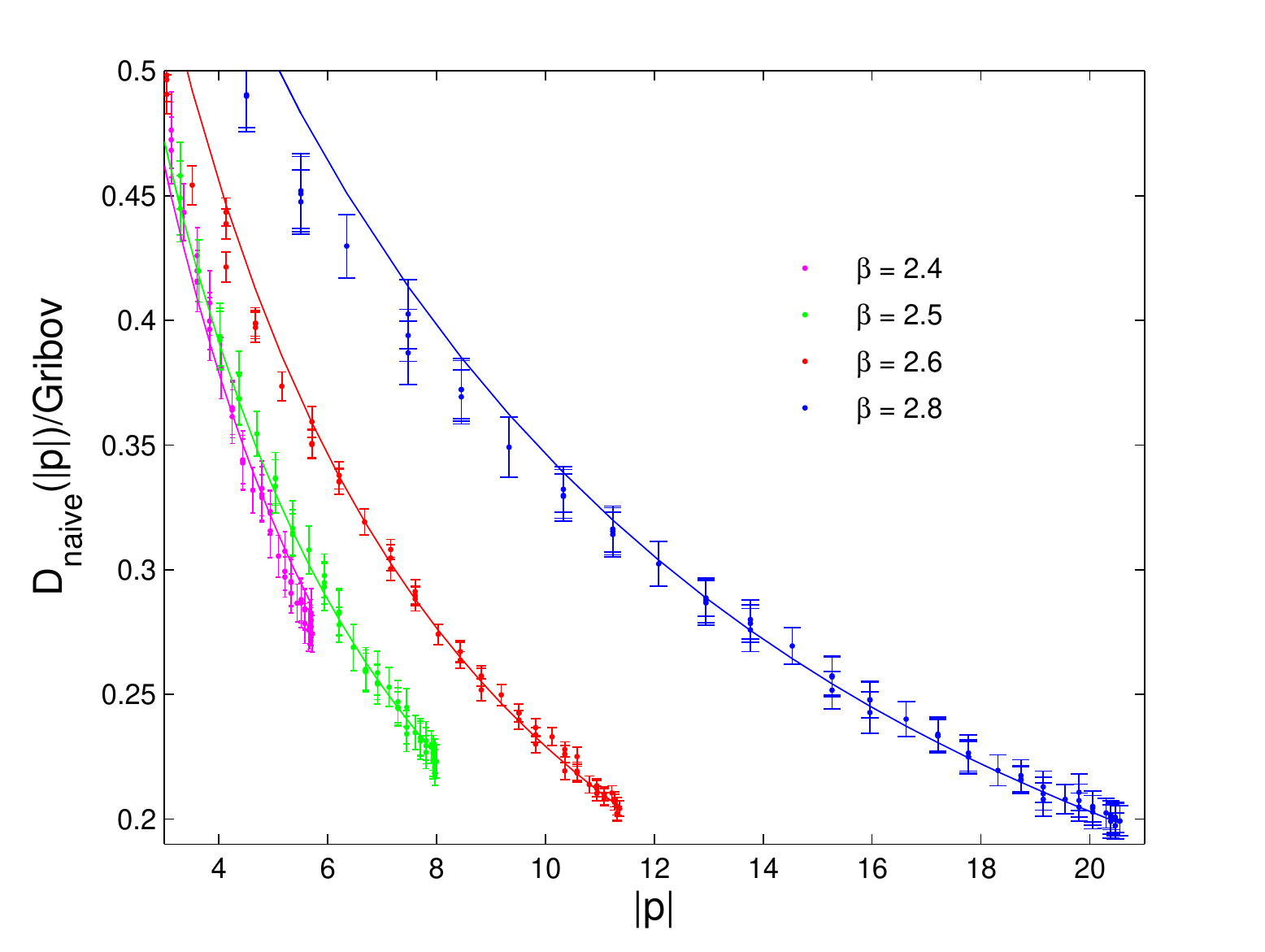}
\caption{Comparison between UV deviations from the Gribov formula in the
MC data \cite{Quandt:2007qd} for the naive static propagator  
$\sum_{p_0} D_\beta(|\vec{p}|,p_0)$ ($\bullet$) and our 
prediction for the leading term 
$B(\frac{4}{4+\hat{p}^2},\frac{1}{2},-\alpha+\frac{1}{2})$ ($\frac{\quad}{\quad}$).}
\label{fig:viol}
\end{figure}

The first observation made in \cite{Burgio:2008jr} is that the lattice 
bare propagator $D_\beta(|\vec{p}|,p_0)$ factorizes as:
\beq
D_\beta(|\vec{p}|,p_0) = \frac{f_\beta(|\vec{p}|)}{|\vec{p}|^2}
\frac{{g}_\beta(z)}{1+z^2}\qquad z=\frac{p_0}{|\vec{p}|}\;.
\label{eq:fact}
\eeq
The denominator $|\vec{p}|^2(1+z^2)$ explicitly accounts for dimensions.
Without loss of generality we can choose $g_\beta(0)= 1$. The data for
${g}_\beta(z)=(1+z^2)D_\beta(|\vec{p}|,p_0) D_\beta(|\vec{p}|,0)^{-1}$ are 
shown
in Fig.~\ref{fig:polaw} for $L=24$. Their leading behaviour
can be well described by a power law $(1+z^2)^{\alpha}$. 
For $\beta \gtrsim 2.3$ the functions $g_\beta$ vary consistently with 
$L$, violating multiplicative renormalizability. 
However as $L\to\infty$ all values of $\alpha$ are compatible with 1
within one or two $\sigma$  i.e. 
$D_\beta(|\vec{p}|,p_0)$ might eventually be $p_0$ independent.
Eq.~(\ref{eq:fact}) has deep consequences on the ``naive'' calculation of 
$D(|\vec{p}|) \propto \sum_{p_0} D(|\vec{p}|,p_0)$ as in 
\cite{Langfeld:2004qs,Quandt:2007qd,Voigt:2007wd}.
Consider different lattice cut-offs for space and time
$\frac{a_s}{a_t} =\xi>1$ and define $\hat{p} = a_s |\vec{p}|$. Neglecting 
subleading terms, for large $L$ we approximate the sum over $p_0$
by an integral yielding:
\bea
D_{\beta}(|\vec{p}|)&\simeq& \int_{-\frac{2}{a_t}}^{\frac{2}{a_t}} \frac{d p_0}{2 \pi}D_\beta(|\vec{p}|,p_0)= 
\frac{f_\beta(|\vec{p}|)}{|\vec{p}|}
\int_{0}^{\frac{2 \xi}{\hat{p}}} \frac{d z}{\pi} 
\left(1+z^2\right)^{\alpha-1} =  \frac{f_\beta(|\vec{p}|)}{|\vec{p}|}
I(\frac{2 \xi}{\hat{p}},\alpha)\;;\nonumber \\
I(\frac{2 \xi}{\hat{p}},\alpha) &=& \frac{1}{2 \pi}
B(\frac{4 \xi^2}{4 \xi^2+\hat{p}^2},\frac{1}{2},-\alpha+\frac{1}{2})\;,
\label{eq:ff}
\eea
where $B(z,a,b)$ is the incomplete beta function. In the lattice 
Hamiltonian limit, corresponding to $\xi\to\infty$ \cite{Burgio:2003in}, 
$I$ becomes $|\vec{p}|$ 
independent\footnote{$I$ can be analytically continued if
$\alpha > \frac{1}{2}$, $\alpha-\frac{1}{2}\not\in \mathbb{N}$.}, 
$\ I(\frac{2 \xi}{\hat{p}},\alpha) \to\frac{1}{2 \sqrt{\pi}} 
\frac{\Gamma(\frac{1}{2}-\alpha)}{\Gamma(1-\alpha)}$. Then 
$D_{\beta}(|\vec{p}|)\propto \frac{{f}_\beta(|\vec{p}|)}{|\vec{p}|}$ 
and multiplicative renormalizability relies solely on $f_\beta(|\vec{p}|)$.
In a standard lattice formulation, however, $\xi\equiv 1$ and 
the extra $|\vec{p}|$ dependence
$B(\frac{4}{4+\hat{p}^2},\frac{1}{2},-\alpha+\frac{1}{2})$ cannot be avoided.
Fig.~\ref{fig:viol} shows, in spite of approximations, nearly perfect 
agreement between Eq.~(\ref{eq:ff}) and the slopes observed in the UV for
the naive definition of $D(|\vec{p}|)$ as in \cite{Quandt:2007qd}. 

\begin{figure}
\includegraphics{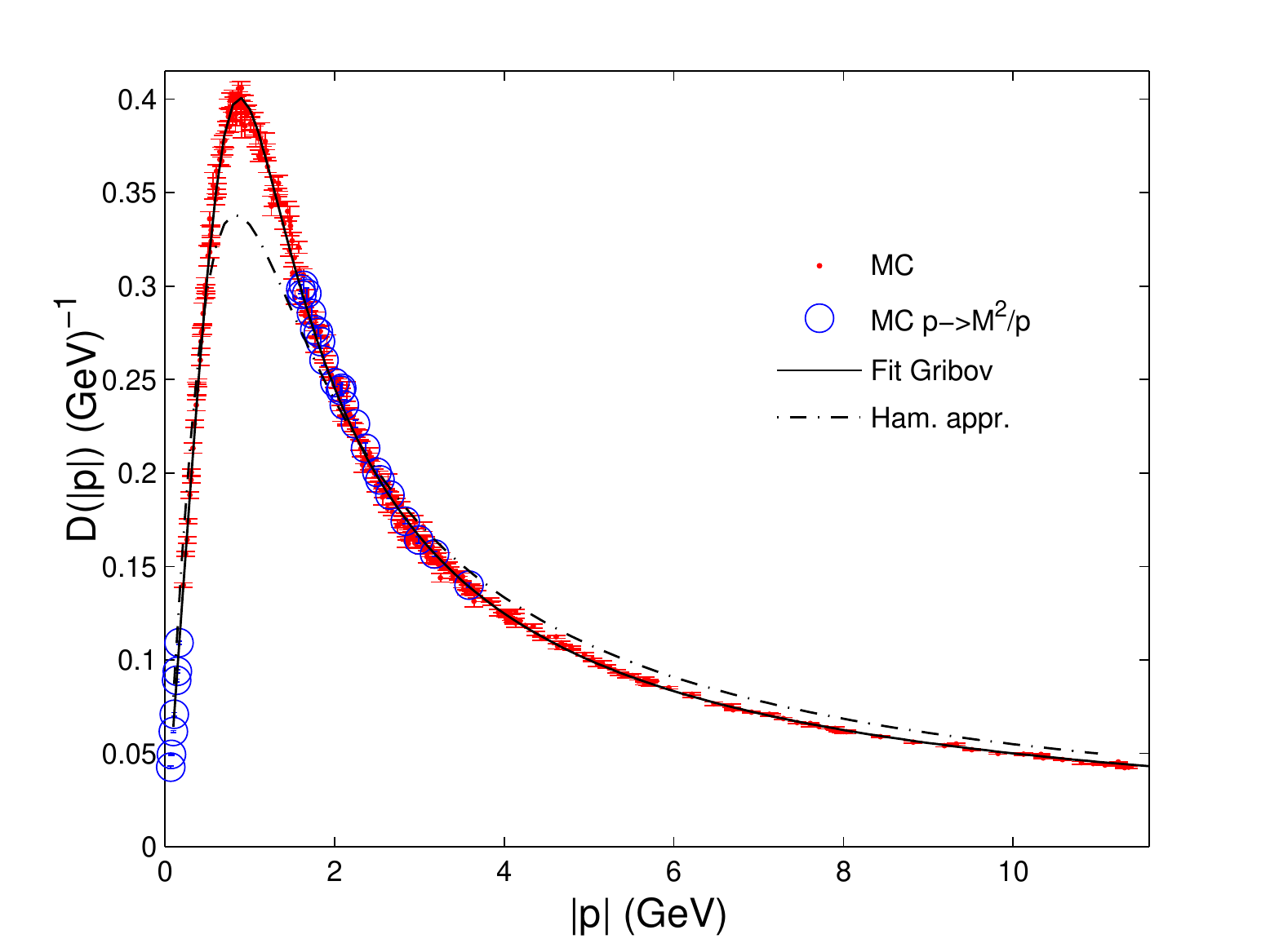}
\caption{The gluon propagator: MC data ($\bullet$), (a few) data for 
$|\vec{p}|\to M^2 |\vec{p}|^{-1}$ ($\bigcirc$), the fit to Gribov's 
formula ($\frac{\quad}{\quad}$) and
the result of the Hamiltonian approach 
\cite{Epple:2006hv} ($\cdot -$).}
\label{fig:glprp}
\end{figure}

The above discussion makes clear that the static propagator should be
defined as 
$D_\beta(|\vec{p}|) =\frac{f_\beta(|\vec{p}|)}{|\vec{p}|}$. To extract it at
available $L$ and $\beta$ we fit ${g}_\beta(z)$ and define:
\beq
f_\beta(|\vec{p}|) = 
D_\beta(|\vec{p}|,p_0)\frac{1+z^2}{{g}_\beta(z)} =: \tilde{D}_\beta
(|\vec{p}|,p_0)
\label{eq:enind}
\eeq 
which is now independent of $p_0$, up to noise.
To improve the signal we average over $p_0$, yielding:
\beq
\tilde{f}_\beta(|\vec{p}|):=\frac{1}{L}\sum_{p_0}\tilde{D}_\beta(
|\vec{p}|,p_0)\;,\quad
D_{\beta}(|\vec{p}|):=\frac{\tilde{f}_\beta(|\vec{p}|)}{|\vec{p}|}\;,
\eeq
Fig.~\ref{fig:glprp} shows the resulting $D_{\beta}(|\vec{p}|)$, which is 
multiplicatively renormalizable.
Fitting
a power law in the IR, $|\vec{p}|^{q} M^{-1-q}$, and a power 
law plus logarithmic corrections in the UV, 
$M^{r-1}|\vec{p}|^{-r}|\log{|\vec{p}|}|^{-s}$ gives
$q = 0.99(1)$, $r = 1.002(3)$, $s=0.002(2)$ and $M=0.88(1)$ GeV, 
with $\chi^2$/d.o.f. all in the range 3.2-3.3, in agreement with the
UV and IR analysis in \cite{Feuchter:2004mk,Epple:2006hv,Schleifenbaum:2006bq}.
We thus constrain 
$q=r=1$, $s=0$ and fit the whole result through the Gribov formula:
\beq
D(|\vec{p}|)=\frac{1}{2 \sqrt{|\vec{p}|^2+\frac{M^4}{|\vec{p}|^2}}}
\eeq
We find just as good agreement ($\chi^2$/d.o.f. $= 3.3$)
again with $M=0.88(1) {\mbox GeV} \simeq 2 \sqrt{\sigma}$.

\section{Conclusions}
\label{sec:conclusions}

We have shown that on the lattice the
static transverse gluon propagator is
multiplicatively renormalizable, IR-UV symmetric and can be well described
by Gribov's formula over the whole momentum range. Its infrared and 
ultraviolet behaviours are in good
agreement with the results obtained in the variational approach to continuum
Yang-Mills theory in Coulomb gauge 
\cite{Feuchter:2004mk,Epple:2006hv,Schleifenbaum:2006bq}. 

\acknowledgments

This work was partly supported by DFG under 
contracts Re856/6-1 and Re856/6-2.


\begin{thebibliography}{99}

\bibitem{Gribov:1977wm}
V.~N.Gribov, \emph{Nucl. Phys} \textbf{{B139}}, 1 (1978).

\bibitem{Zwanziger:1995cv}
{{D.}~{Zwanziger}},
  \emph{Nucl. Phys.} \textbf{{B485}},
  {185} ({1997}), [{\tt hep-th/9603203}].

\bibitem{Szczepaniak:2001rg}
{{A.~P.} {Szczepaniak}}
  {and} {{E.~S.}
  {Swanson}}, \emph{Phys. Rev.}
  \textbf{{D65}}, {025012}
  ({2002}), [{\tt hep-ph/0107078}].

\bibitem{Feuchter:2004mk}
{{C.}~{Feuchter}} {and}
  {{H.}~{Reinhardt}},
  \emph{Phys. Rev.} \textbf{{D70}},
  {105021} ({2004}), [{\tt hep-th/0408236}].

\bibitem{Cucchieri:2000gu}
{{A.}~{Cucchieri}} {and}
  {{D.}~{Zwanziger}},
  \emph{Phys. Rev.} \textbf{{D65}},
  {014001} ({2002}), [{\tt hep-lat/0008026}].

\bibitem{Langfeld:2004qs}
{{K.}~{Langfeld}} {and}
  {{L.}~{Moyaerts}},
  \emph{Phys. Rev.} \textbf{{D70}},
  {074507} ({2004}), [{\tt hep-lat/0406024}].

\bibitem{Quandt:2007qd}
{{M.}~{Quandt}},
  {{G.}~{Burgio}},
  {{S.}~{Chimchinda}},
  {and}
  {{H.}~{Reinhardt}},
  \emph{PoS} \textbf{{LAT2007}},
  {325} ({2007}), [{\tt 0710.0549}].

\bibitem{Voigt:2007wd}
{{A.}~{Voigt}}
{et~al.}
  \emph{PoS} \textbf{{LAT2007}},
  {338} ({2007}), [{\tt 0709.4585}].

\bibitem{Watson:2007mz}
{{P.}~{Watson}} {and}
  {{H.}~{Reinhardt}},
  \emph{Phys. Rev.} \textbf{{D76}},
  {125016} ({2007}), [{\tt 0709.0140}].

\bibitem{Burgio:2008jr}
Burgio, G., Quandt, M. and Reinhardt, H. (2008),
[{\tt 0807.3291}].

\bibitem{Bogolubsky:2005wf}
{{I.~L.} {Bogolubsky}}
  {et~al.}, \emph{Phys. Rev.}
  \textbf{{D74}}, {034503}
  ({2006}), [{\tt hep-lat/0511056}].

\bibitem{Bogolubsky:2007bw}
{{I.~L.} {Bogolubsky}}
  {et~al.}, \emph{Phys. Rev.}
  \textbf{{D77}}, {014504}
  ({2008}), [{\tt 0707.3611}].

\bibitem{Burgio:2003in}
{{G.}~{Burgio}} {et}
{{al.}} ({TrinLat}),
  \emph{Phys. Rev.} \textbf{{D67}},
  {114502} ({2003}), [{\tt hep-lat/0303005}].

\bibitem{Epple:2006hv}
{{D.}~{Epple}},
  {{H.}~{Reinhardt}},
  {and}
  {{W.}~{Schleifenbaum}},
  \emph{Phys. Rev.} \textbf{{D75}},
  {045011} ({2007}), [{\tt hep-th/0612241}].

\bibitem{Schleifenbaum:2006bq}
{{W.}~{Schleifenbaum}},
  {{M.}~{Leder}}, {and}
  {{H.}~{Reinhardt}},
  \emph{Phys. Rev.} \textbf{{D73}},
  {125019} ({2006}), [{\tt hep-th/0605115}].

\end{thebibliography}
\end{document}